\documentstyle[aps,prl,multicol,epsf]{revtex}
\begin{document}
\widetext
\title{Spectral properties of the small polaron}
\author{J. M. Robin}
\address{Max-Planck-Institut f\"ur Physik komplexer Systeme,
Bayreuther Stra{\ss}e 40 Haus 16,
01187 Dresden, Germany}
\date{\today, e-mail: robin@mpipks-dresden.mpg.de}
\maketitle

\begin{abstract}
We compute exactly
both the spectral function of the electron and of the small polaron for the two 
site Holstein model. We find that for intermediary coupling, the 
small polaron is a better fundamental excitation of the system than the
electron. However, the Lang--Firsov approximation fails to predict the right
dispersion relation for the small polaron.
\end{abstract}


\begin{multicols}{2}

\narrowtext

Electronic properties of strongly interacting electron--phonon systems 
are now widely studied since it has been realized experimently 
that the formation of small polarons in some of the new materials, such
as the high $T_{c}$ cuprates\cite{HTC} 
or the perovskite manganates\cite{Millis}, is responsible
in part of some new and unusual physics. 

It is generally believed that the simplest model for the study of small
polarons, is the so called Holstein molecular crystal model\cite{Holstein}.
In the weak coupling limit, electrons behave as good quasiparticles, while
in the strong coupling limit the model is believed to exhibit well 
defined small polarons\cite{Lang-Firsov}.
Perturbation theory towards intermediary coupling can then be used 
from the strong limit fixed point in term of 
small polaron quasiparticles\cite{Lang-Firsov,Stephan}. The validity of
such a perturbation theory rely on the quality of the small polaron 
quasiparticle. If it is destroyed away from infinite coupling then we
can no longer described the physics in term of small polarons and no
longer use perturbation theory.

In this paper, we compute exactly  the spectral functions for the small
polarons in two-site Holstein model for intermediary coupling.
We find that they are good quasiparticles
with a well defined relation dispersion $\omega_{k}^{*}$.
We can then
compute the effective bandwidth of the small polarons $D^{*} = 
max(\omega_{k}^{*}) - min(\omega_{k}^{*})$ and show that it dramatically
decreases with the strength of the electron--phonon coupling constant.
These results clearly show that the physical picture given  by Lang
and Firsov is the right one.

Our calculations are based on a two site
cluster where exact diagonalization are easily performed.
The hamiltonian for the Holstein model is
\[
H = - t \sum_{j,\delta,\sigma} c_{j+\delta,\sigma}^{\dagger} c_{j,\sigma}
+ \omega_{0} \sum_{j} b_{j}^{\dagger} b_{j} 
\]
\vspace{-5mm}
\begin{equation}
- g \omega_{0} \sum_{j,\sigma} c_{j,\sigma}^{\dagger} c_{j,\sigma} 
( b_{j}^{\dagger} + b_{j} ).
\label{bareH}
\end{equation}
The sum over $\delta$ run on the nearest neighbours, $t$ is the hopping 
integral for the tight binding approximation, $\omega_{0}$ is the optical
frequency of the phonon and $g$ is a dimensionless coupling constant.
The polaronic energy is $E_{P}=g^{2}\omega_{0}$ and the usual electron--phonon
coupling constant is $\lambda=E_{P}/zt$ where $z$ is the number of
nearest neighbours ($z=1$ for the two site cluster). 

\begin{figure}
\centerline{\epsfxsize=8cm \epsfbox{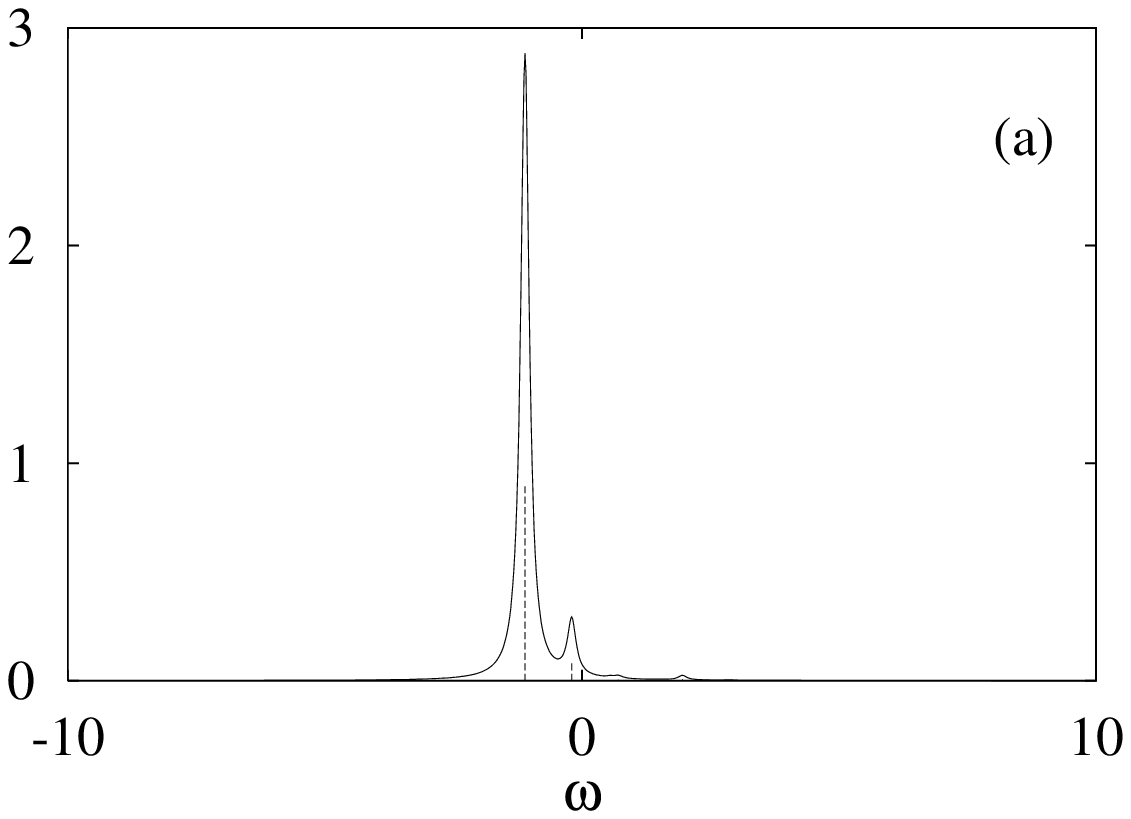}}
\centerline{\epsfxsize=8cm \epsfbox{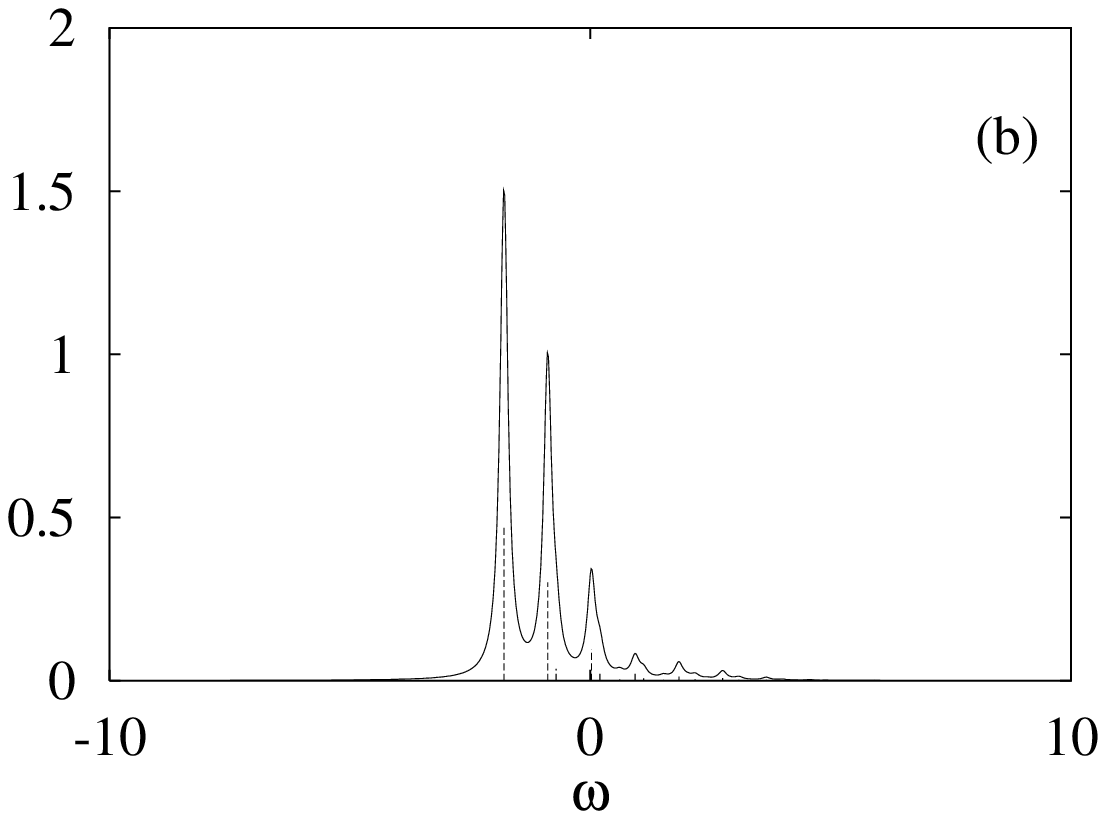}}
\centerline{\epsfxsize=8cm \epsfbox{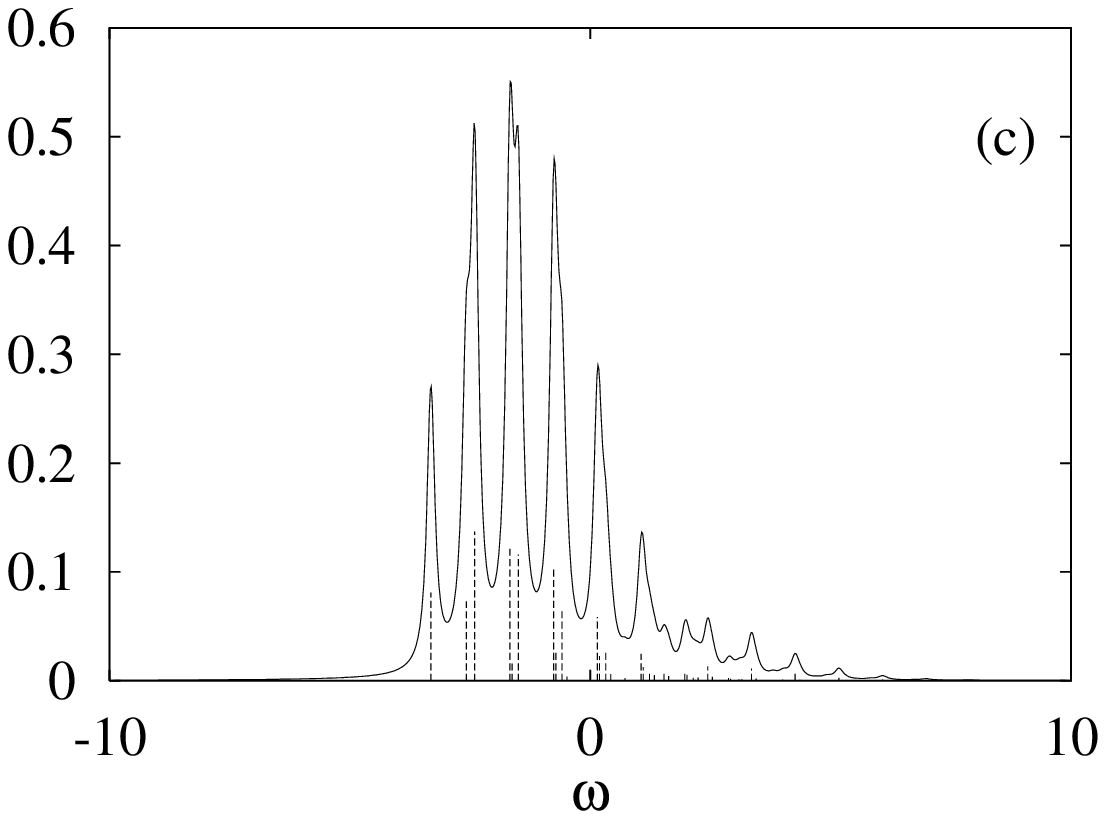}}
\caption{Spectral functions $J(k=0,\omega)$ for the electron, for
three different couplings and $\omega_{0}=1/1.1$, $g=0.3\protect\sqrt{2}$ (a),
$g=0.8\protect\sqrt{2}$ (b), $g=1.3\protect\sqrt{2}$ (c).}
\label{figEL}
\end{figure}

In the atomic limit, $t=0$, the hamiltonian can be diagonalized exactly
by the Lang--Firsov transformation\cite{Mahan}. The new hamiltonian is 
$\tilde{H} = e^{-S} H e^{S}$ with 
$S \; = \; \sum_{j} g (b_{j}^{\dagger} - b_{j}) n_{j}$ 
and $n_{j} = \sum_{\sigma} c_{j,\sigma}^{\dagger}c_{j,\sigma}$. We thus
obtain 
$\tilde{c}_{j,\sigma} = U_{j} c_{j,\sigma}$
and
$\tilde{b}_{j} = b_{j} + g n_{j}$
with
\begin{equation}
U_{j} \; = \; e^{g(b_{j}^{\dagger} - b_{j})}.
\end{equation}
The transformed hamiltonian is 
\[
\tilde{H} = -t \sum_{j,\delta,\sigma} c_{j+\delta,\sigma}^{\dagger} \;
U_{j+\delta}^{\dagger} U_{j} \; c_{j,\sigma} - E_{P} \sum_{j} n_{j}
\]
\vspace{-5mm}
\begin{equation}
+ \omega_{0} \sum_{j} b_{j}^{\dagger} b_{j} 
- 2 E_{P} \sum_{j} n_{j,\uparrow} n_{j,\downarrow}.
\label{transH}
\end{equation}
The hopping term is now a complicated operator. In the Lang--Firsov 
approximation, one obtains an effective hamiltonian for the electrons by
eliminating the phonon states, keeping only the vacuum without phonon. 
At zero temperature one gets $U_{j} \rightarrow e^{-g^{2}/2}$ and thus
an effective hopping integral $t^{*} = e^{-g^{2}} t$ for the small polaron.
In the strong
coupling limit, $g \rightarrow \infty$, this approximation gives localised
polarons, while for finite coupling it gives a polaronic band with the
dispersion relation $\varepsilon^{*}_{\bf k} = -E_{P} + t^{*} \xi_{\bf k}$,
where $\xi_{\bf k}$ is the Fourier transform of the kinetic term. For the
two site cluster, one gets $\varepsilon^{*}_{k} = -E_{P} \pm t^{*}$ for
$k=0,\pi$. If one considers systems with less than two electrons, then 
the interaction term responsible  of the formation of bipolarons (the last
term of eq.(\ref{transH})) vanishes and
the small polarons energies $\varepsilon^{*}_{\bf k}$ are the right excitations
of the system. The spectral function of a small polaron, always in the 
Lang--Firsov approximation, is simply 
$A({\bf k}, \omega) = 2\pi \delta(\omega - \varepsilon^{*}_{\bf k})$.
Inversely, the electrons are no longer good excitations. They carry a cloud
of phonons, with an average number of phonon $ng^{2}$, where $n$ is the
number of electron. The spectral function is given 
by\cite{Mahan,Alex92}
\[
A({\bf k}, \omega) = 
e^{-g^{2}} \;  2 \pi \delta(\omega - \varepsilon^{*}_{\bf k})
\]
\vspace{-5mm}
\begin{equation}
+  \;  e^{-g^{2}} \; \frac{1}{M} \; \sum_{{\bf k}'} \sum_{\ell=1}^{\infty} 
\frac{g^{2\ell}}{\ell!} 2 \pi \delta(\omega - \varepsilon^{*}_{{\bf k}'} - 
\ell \omega_{0}),
\label{AkwLF}
\end{equation}
where $M$ is the number of sites.
Exact diagonalizations of small clusters\cite{ExactDiag} based on the bare 
hamiltonian of eq.(\ref{bareH}) showed that the spectral function for
the electron is too much complicated to be interpreted 
(as can be seen in  fig.\ref{figEL}). For intermediary 
and strong coupling, there is no longer a quasiparticle peak and thus 
impossible to extract a dispersion relation. One possible test for the 
validity of the Lang--Firsov approximation should be to fit the numerical 
results for the spectral function of the electron with the formula of
eq.(\ref{AkwLF}). A more efficient method consist to compute  directly the
spectral function of the small polaron based on the transformed 
hamiltonian of eq.(\ref{transH}). If the small polaron defined by the
canonical transformation is a good quasiparticle
then the spectral function should contains only one main peak with a 
width smaller than the excitation energy. It is then easier to compare
this energy to the one predicts by the Lang--Firsov approximation. 
However the aim of this paper is simply to set up  if the small polaron is
a good quasiparticle for the system.

\begin{figure}
\centerline{\epsfxsize=8cm \epsfbox{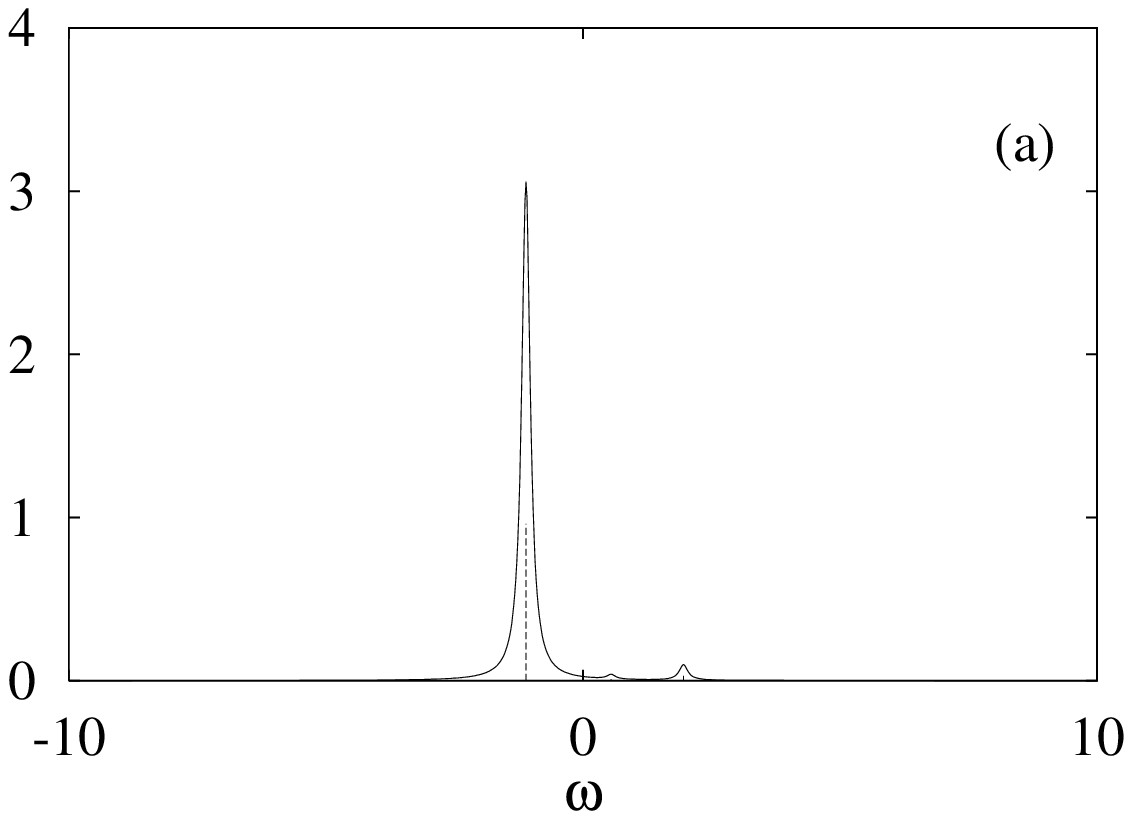}}
\centerline{\epsfxsize=8cm \epsfbox{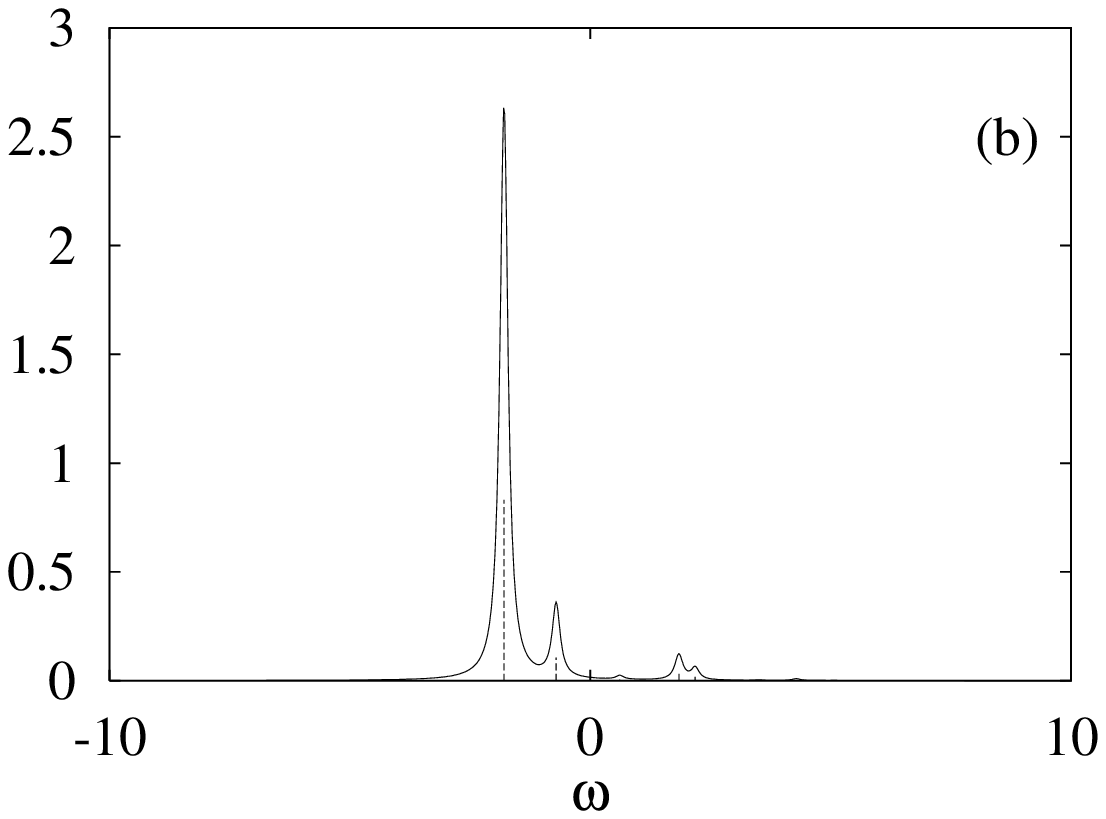}}
\centerline{\epsfxsize=8cm \epsfbox{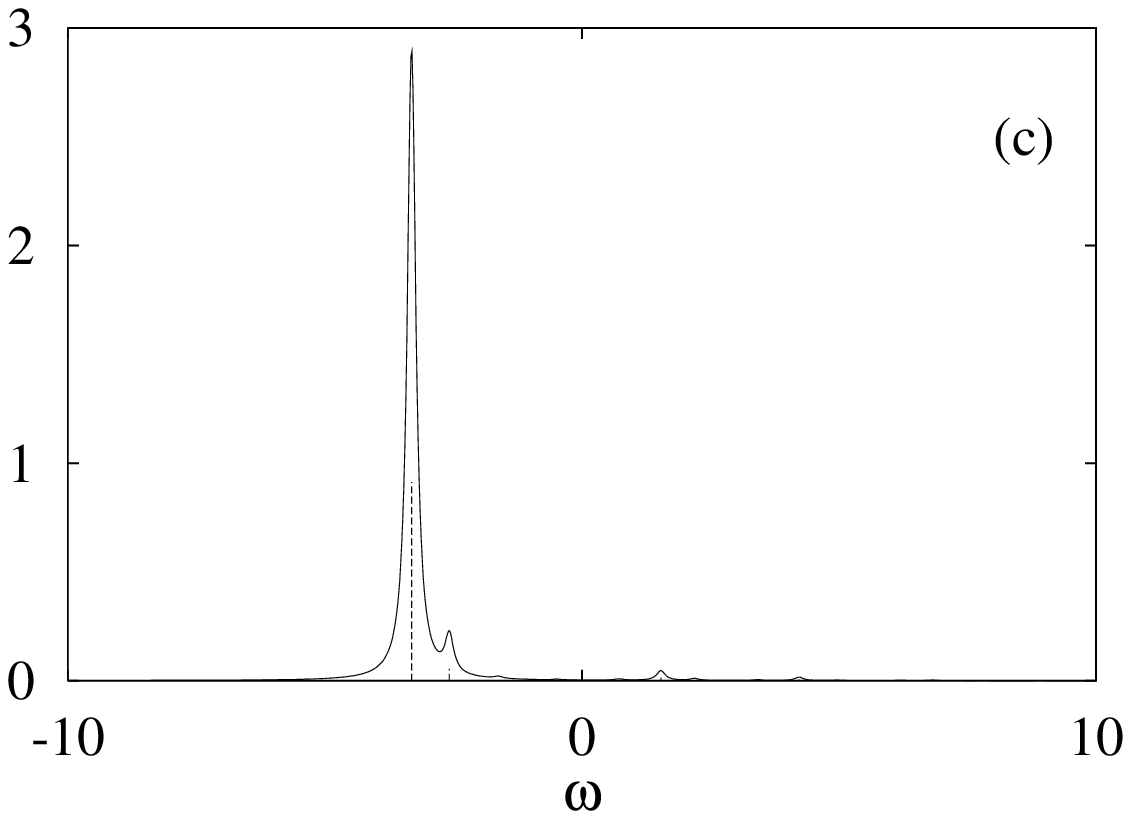}}
\caption{Spectral functions $\tilde{J}(k=0,\omega)$ for the small polaron, for
three different couplings and $\omega_{0}=1/1.1$, $g=0.3\protect\sqrt{2}$ (a),
$g=0.8\protect\sqrt{2}$ (b), $g=1.3\protect\sqrt{2}$ (c).}
\label{figPL}
\end{figure}

In the transformed hamiltonian $\tilde{H}$, the operators $c_{j,\sigma}$ 
and $U_{j}$ commute. One can build separately the operator
$c_{j+\delta,\sigma}^{\dagger}c_{j,\sigma}$ and then work in a subspace
with a fixed number of polarons. The operator $U_{j}$ is computed via its
matrix elements given by
\begin{equation}
U_{n',n} = e^{-g^{2}} \sum_{p=0}^{n} \sum_{p'=0}^{n'} (-)^{p} 
\frac{g^{(p+p')}}{p! p'!} 
\left[\frac{n! n'!}{(n-p)! (n'-p')!}\right]^{1/2}
\end{equation}
and the condition $n-p=n'-p'$. For the phonon basis states, a better 
convergence is obtained if we define some subspaces with constant total number
of phonons\cite{trunc}. We diagonalize matrices with a maximum of $35$ phonons 
(corresponding to $666$ basis states of phonons). Comparison of the
eigenvalues of both the bare and transformed hamiltonians $H$ and $\tilde{H}$
shows that the polaronic representation is better for the results presented
here. In order to deal with the case $\omega_{0} \simeq t$, we
choose the same parameters as in ref.\cite{Rann97} where the spectral 
function $J(k,\omega)$ for the electron has been computed, that is 
$\omega_{0}=1/1.1$,
$g=0.3\sqrt{2}$, $0.8\sqrt{2}$ and $1.3\sqrt{2}$.
We compute the spectral function at zero temperature in the ground state
with zero small polaron defined by
\[
\tilde{J}(k,\omega) = 2 \pi \sum_{m} \left| \langle N=1,m \; | \;
c_{k,\uparrow}^{\dagger} \; | \; N=0, 0 \rangle \right|^{2} \;
\]
\vspace{-5mm}
\begin{equation}
\times \; \delta(\omega + E_{0}^{N=0} - E_{m}^{N=1})
\end{equation}
The number of phonon states has been chosen such that the results 
presented here are well converged. The
criterium is that the transition amplitudes and energies converge 
towards a finite limit as we increase the number of phonon states.
In fig.\ref{figEL}, we show the spectral functions of the electron based
on the hamiltonian of eq.(\ref{bareH}) which correspond with the results
of ref. \cite{Rann97} (we note that these are computed using two different
methods) and in fig.\ref{figPL} we show the spectral functions of the
small polaron based on the hamiltonian of eq.(\ref{transH}). The solid lines
in figures correspond to Lorentzian of width $0.1t$ while peaks 
correspond to the value of the transition amplitude, without the $2\pi$
factor.
We see that for weak enough coupling, both the small polaron and the electron
are good quasiparticle with mainly one transition in the spectral function.
As we increase the coupling we observe for the small polaron that the
spectral weight stays in one peak while it spreads in many peaks more or
less spaced of the energy $\omega_{0}$ for the electron.
For $g=0.3\sqrt{2}$ we find
$\omega_{k=0} \simeq -1.108$ and $\omega_{k=\pi} \simeq 0.977$,
for $g=0.8\sqrt{2}$,
$\omega_{k=0} \simeq -1.795$ and $\omega_{k=\pi} \simeq -1.349$,
and for $g=1.3\sqrt{2}$,
$\omega_{k=0} \simeq -3.314$ and $\omega_{k=\pi} \simeq -3.225$.
All the results are in unit of $t$.
If we compare the renormalized bandwidth $D^{*} = \omega_{k=\pi} -
\omega_{k=0}$ with the Lang--Firsov approximation $D_{LF}^{*} = 2 t
e^{-g^{2}}$ we find a discrepancy of $25\%$ for the two strongest 
couplings.

To conclude, the approach presented here is different from the standard
approach of the so-called polaron problem where one considers a system
with only one electron coupled to the lattice and find the excitation 
energy of the whole Hamiltonian\cite{Feynman}.
Instead, we have shown that the small polaron defined through the
canonical transformation of Lang--Firsov is a good quasiparticle for
strong and intermediary electron-phonon coupling. This method applies
directly to many body problems, such as 
the Hamiltonian of eq.(\ref{transH}) for the Holstein model and for 
arbitrary filling of the system.
Perturbative expansion 
from e.g. the strong coupling 
Lang--Firsov approximation fixed point\cite{Lang-Firsov,Stephan} should be 
relevant.

\acknowledgements
I would like to thank D. Devillers, M. van den Bossche, F.
Vistulo de Abreu,
H. Geoffray  and M. S. Laad for interesting discussions.
Part of
the numerical calculations were performed at Centre de Recherche sur les
Tr\`es Basses Temp\'eratures (CRTBT), Grenoble.

\end{multicols}

\widetext

\end{document}